\documentclass[10pt]{IEEEtran}

\usepackage{amsmath,amssymb,mathtools}    
\usepackage[dvips]{graphicx}   
\usepackage{verbatim}   
\usepackage{color}      
\usepackage{subfigure}  
\usepackage{hyperref}   
\usepackage{epsfig}
\usepackage{float}

\newcommand{\bi}{\begin{itemize}}
\newcommand{\ei}{\end{itemize}}
\newcommand{\beq}{\begin{equation}}
\newcommand{\eeq}{\end{equation}}
\newcommand{\bqn}{\begin{eqnarray*}}
\newcommand{\eqn}{\end{eqnarray*}}
\newcommand{\ba}{\begin{array}}
\newcommand{\ea}{\end{array}}
\newcommand{\bs}{\begin{small}}
\newcommand{\es}{\end{small}}
\newcommand{\nn}{\nonumber}

\providecommand{\norm}[1]{\lVert#1\rVert}

\newcommand{\qed}{\nobreak \ifvmode \relax \else
      \ifdim\lastskip<1.5em \hskip-\lastskip
      \hskip1.5em plus0em minus0.5em \fi \nobreak
      \vrule height0.75em width0.5em depth0.25em\fi}
\DeclarePairedDelimiter\abs{\lvert}{\rvert}

\begin{document}
\title{Hybrid Maximum Likelihood Modulation Classification Using Multiple Radios}
\author{Onur Ozdemir,~\IEEEmembership{Member,~IEEE},
Ruoyu Li, and Pramod K. Varshney,~\IEEEmembership{Fellow,~IEEE}%
\thanks{O. Ozdemir is with ANDRO Computational Solutions, 7902 Turin Road, Rome, NY 13440. R. Li is with the School of EEE, Nanyang Technological University, Singapore. P. K. Varshney is with Department of EECS, Syracuse University, Syracuse, NY 13244. This material is in part based upon work supported by the Government under Contract No. W15P7T-12-C-A040. 
Email: oozdemir@androcs.com, rli11@syr.edu, varshney@syr.edu.}}
\markboth{IEEE COMMUNICATIONS LETTERS (DRAFT)}{}

\date{}

\maketitle

\begin{abstract}
The performance of a modulation classifier is highly sensitive to channel signal-to-noise ratio (SNR). In this paper, we focus on amplitude-phase modulations and propose a modulation classification framework based on centralized data fusion using multiple radios and the hybrid maximum likelihood (ML) approach. In order to alleviate the computational complexity associated with ML estimation, we adopt the Expectation Maximization (EM) algorithm. Due to SNR diversity, the proposed multi-radio framework provides robustness to channel SNR. Numerical results show the superiority of the proposed approach with respect to single radio approaches as well as to modulation classifiers using moments based estimators.
\end{abstract}


\section{Introduction}\label{sec:intro}
Modulation classification (MC) deals with determining the modulation type of a noisy communication signal. It plays an important role in many civilian and military applications, e.g., adaptive cognitive radios for satellite communications \cite{hamkins_06}. A thorough review of MC methods can be found in \cite{su_iet07,su_tsmc_11}. Here, we focus on amplitude-phase modulations and consider the \emph{hybrid maximum likelihood (HML)} approach.
The performance of an MC system using a single radio depends highly on the channel quality, i.e., fading and background noise. In addition, some nuisance parameters, such as signal-to-noise ratio (SNR) and phase offset, are usually unknown which further complicates the classification problem.  Receiver diversity is a common technique used in wireless communication systems to alleviate channel fading effects for demodulation/symbol detection. Similarly, it is natural to argue that using multiple radios for modulation classification, i.e., collaborative MC,  has the potential for improving classification performance compared to a single radio especially in the low to mid signal-to-noise (SNR) regimes. Inspired by this reasoning, collaborative MC approaches have been proposed in \cite{su_milcom04, su_icnsc10, su_sj10, su_milcom11, su_globecom11,su_icc11}. 
Most of these works are based on the distributed detection framework \cite{Varshney:book}, where each radio makes a local (hard or soft) classification decision and then these decisions are fused at a fusion center (FC) to make a global decision  \cite{su_sj10, su_milcom11, su_globecom11}. To the best of our knowledge, there are only two centralized likelihood based approaches proposed in the literature \cite{su_milcom04,su_icc11}. In \cite{su_icc11}, signals from different radios are linearly added to generate a combined signal, which is then used for modulation classification. Linear combining is optimal only if the phase and time information is perfectly known at each radio.  In \cite{su_milcom04}, an antenna array is used to receive the unknown signal. The authors adopt the HLRT framework and use moments based estimators to estimate the unknown signal parameters to simplify the estimation problem. As a result, the estimates in \cite{su_milcom04} are obtained by ignoring the coupling (due to common received constellation symbols) between different antenna elements which results in sub-optimality. 

In this paper, we propose a centralized fusion approach where raw data from local radios as in \cite{su_milcom04,su_icc11} are fused at a fusion center to make the global classification decision. Although the proposed centralized data fusion approach is expected to improve the performance, the resulting MC problem is computationally more complex to solve than a single radio based MC. In order to alleviate this issue, we propose to use the Expectation-Maximization (EM) algorithm \cite{rubin_jrs77}, which significantly simplifies the MC problem along with its nice convergence properties. In an earlier work \cite{silva_tc11}, the EM algorithm was used for the MC problem using a single radio under flat fading channels corrupted by Gaussian mixture noise. Our proposed framework along with the problem formulation for centralized fusion based MC is different from the problem considered in \cite{silva_tc11} even though the EM algorithm is suitable for both. Due to SNR diversity, the proposed centralized data fusion framework significantly improves the MC performance compared to single radio approaches.  Furthermore, our numerical results show that the proposed EM based solution provides superior performance compared to the moments based solution proposed in \cite{su_milcom04} with only a small increase in computational complexity. \vspace{-.25cm}

\section{Problem Formulation}\label{sec:prob}
Consider a radio/sensor network with $L$ sensors observing the same communication signal with a block of $N$ constellation (information) symbols that undergo flat block fading. These sensors are located more than half wavelength apart so that they experience independent fading. We assume that timing and frequency offsets have been perfectly estimated and the pulse-shaping filter is known. Under these assumptions, the received baseband observation sequence at sensor $l$ is
\beq
r_{l,n} = a_l e^{j\theta_l} I_n + w_n , \label{eq:simp_rec}
\eeq
where $l=1,\ldots,L$, $n=0,\ldots,N-1$, $I_n$ is the $n^{th}$ complex constellation symbol of the block, $w_n$ is the additive complex zero-mean white Gaussian noise with variance $N_0$, and $a_l$ and $\theta_l$ are the channel gain and the channel phase at sensor $l$, respectively. 
In this model, $\{a_l\}_{l=1}^L$, $\{\theta_l\}_{l=1}^L$, $\{I_n\}_{n=0}^{N-1}$, $N_0$ are the unknown model parameters. The unknown parameter vector can be expressed as $\tilde{\mathbf{u}}\triangleq\left[\mathbf{a},\mathbf{\theta},\mathbf{I},N_0\right]$, where $\mathbf{a}\triangleq\left[a_1,\ldots,a_L\right]^T$, $\mathbf{\theta}\triangleq\left[\theta_1,\ldots,\theta_L\right]^T$ and $\mathbf{I}\triangleq\left[I_0,\ldots,I_{N-1}\right]^T$\footnote{Superscript T denotes vector/matrix transpose.}. We assume that noise is independent across sensors. Suppose there are $S$ candidate modulation formats under consideration and let  $I_n^{(i)}$ denote the constellation symbol at time $n$ corresponding to modulation $i\in\{1,\ldots,S\}$. We assume that \textit{a priori} probabilities of the modulation formats are identical, in which case the optimal Bayesian classifier takes the form of a maximum likelihood (ML) classifier. In the hybrid maximum likelihood approach \cite{su_iet07}, the LF is marginalized over the unknown constellation symbols $I_n$ and then maximized over the remaining unknown (nuisance) parameters. Let $\mathbf{r}$ denote the observation vector defined as $\mathbf{r} \triangleq [\mathbf{r}_1^T,\ldots,\mathbf{r}^T_L]^T$ where  $\mathbf{r}_l \triangleq [r_{l,0},\ldots,r_{l,N-1}]^T$ and $H_i$ represent the hypothesis associated with modulation format $i$. Let $\mathbf{u}\triangleq\left[\mathbf{a},\mathbf{\theta},N_0\right]$ and $p_i(\mathbf{r}|\mathbf{u})\triangleq p(\mathbf{r}|H_i,\mathbf{u})$\footnote{Throughout the paper, we use the notation $p_i(\cdot)$ to denote $p(\cdot|H_i)$.} denote the conditional probability density function (pdf) of $\mathbf{r}$ conditioned on the unknown modulation format $i$ and the unknown parameter vector $\mathbf{u}$.
%
%
Given $I_n$ and hypothesis $H_i$, we have the following
\beq
p_i(r_{1,n},\ldots,r_{L,n}|I_n,\mathbf{u}) = \prod_{l=1}^L p_i(r_{l,n}|I_n,\mathbf{u}).
\eeq
After marginalizing over $I_n$ and using the fact that noise is independent across samples, we get
\beq
p_i(\mathbf{r}|\mathbf{u}) =\frac{1}{M_i^N}\prod_{n=1}^N\sum_{m=1}^{M_i}\prod_{l=1}^L p_i(r_{l,n}|I_n^{m,(i)},\mathbf{u}), \label{eq:av2}
\eeq
where $p_i(r_{l,n}|I_n^{m,(i)},\mathbf{u})$ denotes the pdf of a complex Gaussian distribution with mean $a_l e^{j\theta_l} I_n^{m,(i)}$ and variance $N_0$, and $M_i$ and $I_n^{m,(i)}$ are the number of constellation symbols and the $m^{th}$ constellation symbol in modulation $i$, respectively. Note that, in (\ref{eq:av2}), the constellation symbols are assumed to have equal \emph{a priori} probabilities, i.e., $p(I_n^{m,(i)}|H_i)=1/M_i$. Without loss of generality, we further assume that $\mathbb{E}\{\abs{I_n^{(i)}}^2\}=1$, where $\mathbb{E}\{\cdot\}$ denotes statistical expectation. By using (\ref{eq:av2}) and the fact that each constellation symbol is independent, we can discard the irrelevant terms and obtain the log-likelihood function shown in (\ref{eq:llf1}) on the top of next page.
%
In the HML approach, the modulation that maximizes the resulting LLF is selected as the final decision, i.e., $\hat{i} = \arg\max_i \Lambda_i(\hat{\mathbf{u}}_i)$, where
\begin{figure*}
\beq
\Lambda_i(\mathbf{u}) = -N\ln M_i - LN \ln N_0 +  \sum_{n=0}^{N-1}\ln \left(\sum_{m=1}^{M_i}
\exp\left(-\frac{1}{N_0}\sum\limits_{l=1}^L \abs[\Big]{r_{l,n}- a_l e^{j\theta_l} I_n^{m,(i)}}^2 \right)\right) \label{eq:llf1}
\eeq\vspace{-.25in}
\underline{\hspace{\textwidth}}
\end{figure*}
\beq
\hat{\mathbf{u}}_i = \arg\max_\mathbf{u} \Lambda_i(\mathbf{u}).\label{eq:ml2}\eeq
From (\ref{eq:llf1}), we can observe that the problem of finding the global maximum of $\Lambda_i(\mathbf{u})$ with respect to $\mathbf{u}$ is a $2L+1$ dimensional non-convex optimization problem which is prohibitively complex to solve in general. Furthermore, there is coupling between the unknowns of different sensors due to common unknown constellation symbols, i.e., the problem cannot be decoupled across sensors into multiple lower dimensional optimization problems. There is no closed-form analytical solution. Therefore, either numerical methods or approximation techniques need to be employed. In the following section, we discuss our approach for solving this problem which is based on the Expectation-Maximization (EM) algorithm.\vspace{-.2cm}

\section{The EM Algorithm}\label{sec:em}
Suppose for now that the modulation $i$ is under consideration and the constellation symbol vector $\mathbf{I}$ is known. In this case, we have the following closed-form expressions for the ML estimators
\beq
    \hat{\theta}_l = \tan^{-1}\left(\Im(\mathbf{I}^H\mathbf{r}_l) / \Re(\mathbf{I}^H\mathbf{r}_l)\right),
\eeq
\beq
\hat{a}_l = \Re\left(e^{-j\hat{\theta}_l} \mathbf{I}^H \mathbf{r}_l\right)/\norm{\mathbf{I}}^2,
\eeq
\beq
\hat{N}_0 = \frac{1}{LN} \sum_{n=0}^{N-1} \sum_{l=1}^L \abs[\big]{r_{l,n} - \hat{a}_l e^{j \hat{\theta}_l} I_n}^2,
\eeq
where $\Re(\cdot)$ and $\Im(\cdot)$ denote the real and imaginary parts of a complex number, respectively, and $H$ denotes the Hermitian of a complex vector/matrix. From the above closed-form expressions, it is clear that when $\mathbf{I}$ is known, the maximization problem (for estimating $a_l$ and $\theta_l$) decouples between different sensors. Due to the fact that the ML estimation problem is significantly simpler when the constellation symbols are known, we adopt the well-known EM algorithm \cite{rubin_jrs77} to solve this problem by treating constellation symbols as missing (unobserved) data. The EM algorithm is an iterative method which enables the computation of ML estimates, especially well suited to problems where ML estimation is intractable due to the presence of missing data. In our case, the constellation symbols represent missing data. We can formally describe the EM algorithm for our problem in (\ref{eq:ml2}) as follows \cite{rubin_jrs77}. Let us define the so-called \emph{complete data} $\mathbf{x}\triangleq[\mathbf{r}^T,\mathbf{I}^T]$. The EM algorithms starts from an initial estimate $\hat{\mathbf{u}}_i^{(0)}$ and performs the following two steps at iteration $t+1$: the expectation step (E-step) and the maximization step (M-step) given as 
\begin{align}
&\text{\textbf{E-step:}}\quad Q(\mathbf{u}_i|\hat{\mathbf{u}}_i^{(t)})=\mathbb{E}\left\{\ln p_i(\mathbf{x}|\mathbf{u}_i)|\mathbf{r},\hat{\mathbf{u}}_i^{(t)}\right\},\label{eq:estep}\\
&\text{\textbf{M-step:}}\quad \hat{\mathbf{u}}_i^{(t+1)}=\arg\max_{\mathbf{u}_i} Q(\mathbf{u}_i|\hat{\mathbf{u}}_i^{(t)}).\label{eq:mstep}
\end{align}
Given the fact that the unknown parameter vector $\mathbf{u}$ is independent of the transmitted constellation symbols $\mathbf{I}$, the E-step in (\ref{eq:estep}) reduces to
\beq
Q(\mathbf{u}_i|\hat{\mathbf{u}}_i^{(t)}) = \sum_{\mathbf{I}} \ln p_i(\mathbf{r}|\mathbf{I},\mathbf{u}_i) P_i\left(\mathbf{I}|\mathbf{r},\hat{\mathbf{u}}_i^{(t)}\right). \label{eq:estep2}
\eeq
We define  $\mathbf{r}_n\triangleq[r_{1,n},\ldots,r_{L,n}]^T$. Let $\alpha_n^{m,(t)}\triangleq P_i\left(I_n=I^{m}|\mathbf{r}_n,\hat{\mathbf{u}}_i^{(t)}\right)$, $m\in\{1,\ldots,M_i\}$, denote the \emph{a posteriori} probability of the unknown constellation symbol which can be calculated as

\begin{small}
\begin{align}
\alpha_n^{m,(t)}\triangleq P_i\left(I_n=I^{m}|\mathbf{r}_n,\hat{\mathbf{u}}_i^{(t)}\right) & = \frac{p_i\left(I_n=I^{m}, \mathbf{r}_n|\hat{\mathbf{u}}_i^{(t)}\right)}{P_i\left(r_n|\hat{\mathbf{u}}_i^{(t)}\right)} \nn\\
&\stackrel{(a)}{=}\frac{p_i\left(\mathbf{r}_n|I_n=I^{m},\hat{\mathbf{u}}_i^{(t)}\right)}{\sum\limits_{k=1}^{M_i} p_i\left(\mathbf{r}_n|I_n=I^{k},\hat{\mathbf{u}}_i^{(t)}\right)} \nn
\end{align}
\beq
= \frac{\exp\left(-\sum\limits_{l=1}^L\abs{r_{l,n}^{(t)}- \hat{a}_l^{(t)} e^{j \hat{\theta}_l^{(t)}} I^m}^2 / \hat{N}_0^{(t)}\right)}
{\sum\limits_{k=1}^{M_i} \exp\left(-\sum\limits_{l=1}^L\abs{r_{l,n}^{(t)}- \hat{a}_l^{(t)} e^{j \hat{\theta}_l^{(t)}} I^k}^2 / \hat{N}_0^{(t)}\right)}.
\label{eq:post}
\eeq\end{small}While deriving \eqref{eq:post} in step $(a)$, we have used the assumption that $P_i\left(I_n=I^{m}|\hat{\mathbf{u}}_i^{(t)}\right)=1/M_i$, $m\in\{1,\ldots,M_i\}$. Let us also define
\beq
v_n^{(t)}\triangleq\sum\limits_{m=1}^{M_i} \alpha_n^{m,(t)} I^m,\quad E^{(t)} \triangleq \sum\limits_{n=0}^{N-1}\sum\limits_{m=1}^{M_i}\alpha_n^{m,(t)} \abs[\big]{I_n^{m}}^2. \label{eq:exp}
\eeq
Note that $v_n^{(t)}$ and $E^{(t)}$ represent the \emph{a posteriori} expectations of the constellation symbol at time $n$ and the normalized energy of the transmitted discrete-time signal, respectively. Substituting (\ref{eq:post})-(\ref{eq:exp}) in (\ref{eq:estep2}) and carrying out the maximization in (\ref{eq:mstep}) by taking the first derivatives and setting them to zero, we obtain the following closed-form expressions for the $(t+1)$-th step in the EM algorithm
\beq
    \hat{\theta}_l^{(t+1)} = \tan^{-1}\left(\Im(\mathbf{\Upsilon}^{(t)^H}\mathbf{r}_l) / \Re(\mathbf{\Upsilon}^{(t)^H}\mathbf{r}_l)\right),\label{eq:it1}   
\eeq
\beq
\hat{a}_l^{(t+1)}= \Re\left(e^{-j\hat{\theta}_l^{(t+1)}} \mathbf{\Upsilon}^{(t)^H} \mathbf{r}_l\right) / E^{(t)},
\eeq
\beq
\hat{N}_0^{(t+1)} = \frac{1}{LN} \sum_{n=0}^{N-1} \sum\limits_{m=1}^{M_i} \alpha_n^{m,(t)} \sum_{l=1}^L \abs[\Big]{r_{l,n} - \hat{a}_l^{(t+1)} e^{j \hat{\theta}_l^{(t+1)}} I^{m}}^2,\label{eq:it3}
\eeq
where $\mathbf{\Upsilon}^{(t)} \triangleq [\upsilon^{(t)}_0,\ldots,\upsilon^{(t)}_{N-1}]^T$. Note from \eqref{eq:post} that the EM algorithm uses information from all the sensors to update posterior probabilities of constellation symbols. This is the crucial step to enable data fusion. After this step, the estimation process becomes decoupled among sensors as shown in \eqref{eq:it1}-\eqref{eq:it3}, which significantly simplifies the original coupled ML estimation problem. One important property of the EM algorithm is that the original LLF monotonically increases at every iteration and converges to a stationary point \cite{wu_ann83}. However, this stationary point can be a local maxima, therefore, either a good initialization or multiple initializations are needed to guarantee convergence to a good stationary point. 

\section{EM Initialization}
There are many methods to initialize the EM algorithm. One method is to use simple blind estimators. In the MC literature, there have been attempts to use simple estimators due to the complexity associated with ML estimators \cite{su_milcom04,dobre_twc09}. These estimators are based on the method of moments (MoM). More specifically the authors in \cite{su_milcom04,dobre_twc09} adopt the second and fourth order moments ($M_2M_4$) parameter estimators \cite{pauluzzi_tc00} for the MC problem. The $M_2M_4$ estimators for $a_l$ and $N_0$ are given, respectively, as $\hat{a}_{l,(i)} = \left(\frac{2\hat{M}_{2,l}^2-\hat{M}_{4,l}}{2-\mathbb{E}\{\abs{I^{(i)}}^4\}}\right)^{1/4}$, $\hat{N}_{0_{(i)}} = \sum_{l=1}^L \hat{N}_{0_{l,(i)}}$,
where $\hat{N}_{0_{l,(i)}} = \hat{M}_{2,l} - \hat{a}_{l,(i)}^2$, 
 $\hat{M}_{2,l}=N^{-1}\sum_{n=0}^{N-1}\abs{r_{l,n}}^2$ and $\hat{M}_{4,l}=N^{-1}\sum_{n=0}^{N-1}\abs{r_{l,n}}^4$. Regarding phase initialization, the MoM estimators depend on the modulation format under consideration. A common MoM phase estimator is the $K$th power estimator for the general $2\pi/K$-rotationally symmetric constellations given as $\theta_{l,(i)} = K^{-1}\arg\left(\mathbb{E}\{I_n^{*K}\}\sum_{n=0}^{N-1}r_{l,n}^{K}\right)$. For M-PSK, $K=M$ whereas for M-QAM $K=4$. In \cite{su_milcom04,dobre_twc09}, special cases of these estimators have been used for the MC problem. It was shown in \cite{jonghe_tc94} that the $K$th power phase estimator is equivalent to the ML estimator in the limit as SNR$\rightarrow0$. For the special case of cross QAM modulations (e.g. 32-QAM), another blind  estimator based on the eighth order moments has been proposed in \cite{cartwright_spl01}, which provides improved performance over the $K$th power estimator for cross QAM modulations.

Other methods to initialize the EM algorithm include performing a coarse grid search over the parameter space or using a stochastic optimization algorithm such as simulated annealing (SA) \cite{bertsemas_ss93}. It is also possible to use hybrid approaches such as the following. When SNR is small, the $K$th power phase estimator can be used to initialize $\hat{\theta}$ due to the result in \cite{jonghe_tc94}. However, when SNR is moderate or high, a coarse grid search or SA could work better for phase initialization combined with $M_2M_4$ estimator for $\hat{a}$ and $\hat{N}_0$ initialization. It should be noted that in some cases $M_2M_4$ estimator could result in negative or imaginary $\hat{N}_0$ which is inaccurate. In these cases, we could average only the $\hat{N}_{0_l}$s that are positive or we could use a coarse grid search if all $\hat{N}_{0_l}$s are inaccurate.\vspace{-.2cm} 

\section{Numerical Results}
\begin{figure}
\begin{center}
\includegraphics[width=0.32\textwidth,height=!]{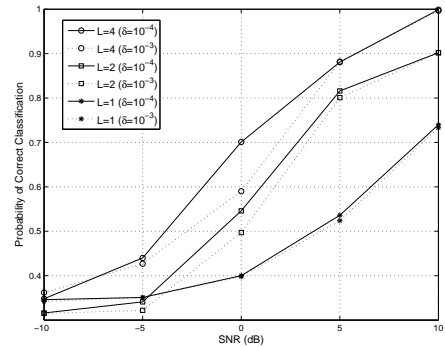}
\end{center}
\vspace{-.2in}
\caption{\small $P_c$ versus SNR for different stopping criteria}\label{fig:pc_vs_snr}\vspace{-.15in}
\end{figure}

We consider a ternary MC scenario where the modulations under consideration are 16-QAM, 32-QAM and 64-QAM. Each channel is modeled as a Rayleigh block fading channel, i.e., $a_l$ is a Rayleigh distributed random variable with scale parameter $\sigma$. The average channel SNR is given as $\mathbb{E}\{a_l^2\}/N_0=2\sigma^2/N_0$. Channel phase $\theta$ is uniformly distributed in $[-\pi,\pi)$. We fix $2\sigma^2=1$ and vary the noise power $N_0$ to simulate different channel SNRs. Each radio experiences independent fading (with identical statistics) resulting in SNR diversity among radios. Since EM is an iterative algorithm, we continue the iterations until the relative improvement of the likelihood function is within a stopping criterion $\delta$. We use $M_2M_4$ estimators to initialize $\hat{a}_l$ and $\hat{N}_0$. As for $\hat{\theta}$ initialization, we use the $4$th power estimate \cite{jonghe_tc94} for 16- and 64-QAMs, and the eight order moment based estimate  \cite{cartwright_spl01} for 32-QAM. For SNR $\geq 10$ dB, we also perform a coarse grid search around the initial phase estimate since MoM phase estimates get farther away from the ML estimates as SNR increases. Fig. \ref{fig:pc_vs_snr} shows the average probability of correct classification ($P_c$) for all the modulations under consideration versus channel SNR under different number of radios for $\delta=10^{-4}$ and $10^{-3}$. Low to mid channel SNR regimes are considered since this is where the multi-radio approach is expected to provide significant performance improvement. The number of samples is fixed at $N=500$ and each $P_c$ is based on $1000$ Monte Carlo runs. Note that the channel gains also vary across different Monte Carlo runs. It is clear from Fig. \ref{fig:pc_vs_snr} that a centralized data fusion based multi-radio approach is the key to improving performance at low to mid SNR regimes. For example, when SNR = 5 dB, we can increase $P_c$ from around $0.54$ up to $0.88$ with four radios compared to a single radio. The trade-off is the cost of the radios, increased bandwidth requirement and synchronization overhead that is needed between radios. 

In our simulations, we observed that for $\delta<10^{-4}$ there is no significant improvement in the performance, and for $\delta>10^{-3}$ the performance degrades significantly. Tables \ref{tab1} and \ref{tab2} present the average number of iterations (rounded to the nearest integer) and $P_c$ for SNR = 0 dB and 5 dB, respectively. We can observe from the tables that as the number of sensors increases, more iterations are needed for convergence. Similarly, higher SNR values require more iterations. This is due to the fact that blind phase estimates used for initialization are very close to the ML estimates for smaller SNR values whereas this is not the case for larger SNRs. 

\begin{figure}
\begin{center}
\includegraphics[width=0.32\textwidth,height=!]{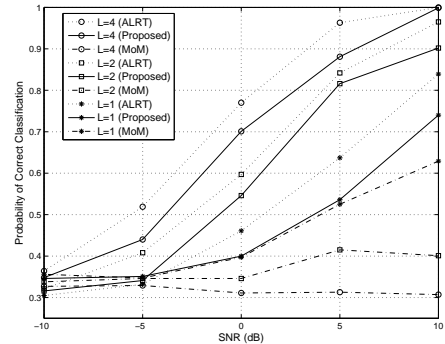}
\end{center}
\vspace{-.2in}
\caption{\small Comparison of different classifiers}\label{fig:pc_vs_n}\vspace{-.25in}
\end{figure}

Fig. \ref{fig:pc_vs_n} shows comparison of the proposed EM based classifier with the clairvoyant classifier (ALRT) \cite{mendel_tc_00}, which assumes that the SNRs and phases are perfectly known. The ALRT serves as an upper performance bound. It is clear from the figure that the proposed classifier performs close to this upper performance bound. For comparison, we also include the results obtained by only using the MoM estimates, i.e., initial points for the EM algorithm, similar to \cite{su_milcom04}. We observe from the figure that the proposed EM based approach provides superior classification performance compared to the MoM based approach. In fact, it is surprising to see that the performance of classifiers using MoM based estimation degrades as the number of radios increases, to the point where they are no better than simple guessing. This is due to the fact that MoM estimators do not necessarily maximize the LF and they do not take into account coupling between estimates of different radio signals due to common constellation symbols. These factors result in poor sub-optimality of MoM based modulation classifiers when multiple sensors are used. \vspace{-.2cm}

\begin{table}
\caption{Effect of Stopping Criterion (SNR$=0$ dB)}\vspace{-.5cm}
\label{tab1}
\begin{center}
\begin{tabular}{|c|c|c|c|c|c|c|}
  \hline
  &\multicolumn{2}{|c|}{L=1}&\multicolumn{2}{|c|}{L=2}&\multicolumn{2}{|c|}{L=4} \\
  \hline
Stop. Criterion & Iter. & $P_c$ & Iter. & $P_c$ & Iter. & $P_c$\\
 \hline
 $\delta=10^{-4}$& 5 & 0.4 & 28 & 0.546 & 33 & 0.701\\
 \hline
 $\delta=10^{-3}$& 4 & 0.398 & 15 & 0.497 & 15 & 0.59\\
\hline
\end{tabular}\vspace{-.5cm}
\end{center}
\end{table}
\begin{table}
\caption{Effect of Stopping Criterion (SNR$=5$ dB)}\vspace{-.5cm}
\label{tab2}
\begin{center}
\begin{tabular}{|c|c|c|c|c|c|c|}
  \hline
  &\multicolumn{2}{|c|}{L=1}&\multicolumn{2}{|c|}{L=2}&\multicolumn{2}{|c|}{L=4} \\
  \hline
Stop. Criterion & Iter. & $P_c$ & Iter. & $P_c$ & Iter. & $P_c$\\
 \hline
 $\delta=10^{-4}$& 20 & 0.536 & 50 & 0.816 & 79 & 0.881\\
 \hline
 $\delta=10^{-3}$& 5 & 0.524 & 20 & 0.801 & 33 & 0.882\\
\hline
\end{tabular}\vspace{-.7cm}
\end{center}
\end{table}


\bibliographystyle{IEEEtran}
\bibliography{Book,Conf,Journal}



\end{document}